\documentclass{emulateapj}
\usepackage{graphicx}
\usepackage{amssymb,amsfonts,amsmath,amstext,amsgen,amsopn,amsxtra,indentfirst,times}

\begin{document}

\title{Atmospheric Chemistry for Astrophysicists: \\ A Self-consistent Formalism and Analytical Solutions for Arbitrary C/O}

\author{Kevin Heng\altaffilmark{1}}
\author{James R. Lyons\altaffilmark{2}}
\author{Shang-Min Tsai\altaffilmark{1}}
\altaffiltext{1}{University of Bern, Center for Space and Habitability, Sidlerstrasse 5, CH-3012, Bern, Switzerland.  Email: kevin.heng@csh.unibe.ch}
\altaffiltext{2}{Arizona State University, School of Earth and Space Exploration, Bateman Physical Sciences, Tempe, AZ 85287-1404, U.S.A.}

\begin{abstract}
We present a self-consistent formalism for computing and understanding the atmospheric chemistry of exoplanets from the viewpoint of an astrophysicist.  Starting from the first law of thermodynamics, we demonstrate that the van't Hoff equation (which describes the equilibrium constant), Arrhenius equation (which describes the rate coefficients) and procedures associated with the Gibbs free energy (minimisation, rescaling) have a common physical and mathematical origin.  We address an ambiguity associated with the equilibrium constant, which is used to relate the forward and reverse rate coefficients, and restate its two definitions.  By necessity, one of the equilibrium constants must be dimensionless and equate to an exponential function involving the Gibbs free energy, while the other is a ratio of rate coefficients and must therefore possess physical units.  We demonstrate that the Arrhenius equation takes on a functional form that is more general than previously stated without recourse to tagging on ad hoc functional forms.  Finally, we derive analytical models of chemical systems, in equilibrium, with carbon, hydrogen and oxygen.  We include acetylene and are able to reproduce several key trends, versus temperature and carbon-to-oxygen ratio, published in the literature.  The rich variety of behavior that mixing ratios exhibit as a function of the carbon-to-oxygen ratio is merely the outcome of stoichiometric book-keeping and not the direct consequence of temperature or pressure variations.
\end{abstract}


\section{Introduction}

\subsection{Preamble}

Understanding chemistry is indispensible to deciphering the abundances of atomic and molecular species present in an exoplanetary atmosphere.  Despite its somewhat late start in the study of exoplanets (e.g., \citealt{bs99,zahnle09,moses11,moses13a,moses13b,hu12,hu13,madhu12,ly13,blecic15,venot15}), atmospheric chemistry has a long and rich history in the Earth and planetary sciences and the study of brown dwarfs (e.g., \citealt{prinn77,barshay78,allen81,fegley96,lodders02,ciesla06}).  Yet, a first-principles, self-consistent formalism that unifies all of the quantities and terminology in a form that is useful for astrophysicists is missing from the literature.  For example, there is more than one definition of the ``equilibrium constant".  

Within the same framework, we demonstrate that the van't Hoff equation (which describes the dimensionless form of the equilibrium constant), the Arrhenius equation (which describes the rate coefficients) and procedures associated with the Gibbs free energy (minimisation and scaling) all originate from the first law of thermodynamics.  The foundations of atmospheric chemistry are built upon statistical mechanics, since the first law derives from it.  To demonstrate the usefulness of our formalism, we use it to compute analytical solutions of chemical systems with pure hydrogen and with carbon, oxygen and hydrogen (gas phase only); we show that these solutions generalise the work of \cite{bs99} and correctly reproduce all of the expected trends.

\subsection{Survey of monographs}

The novelty of the present study is not in the individual formulae stated, which are mostly previously known; with one exception, we certainly do not claim to be the first to derive these formulae.  Rather, it is in the way these results are derived and unified under a common, self-consistent, mathematical formalism that is accessible and palatable to astrophysicists (rather than to chemists).  We will now demonstrate this claim of novelty by surveying several textbooks in chemistry.

Specifically, equations (\ref{eq:gibbs_differential}), (\ref{eq:gibbs_total}), (\ref{eq:gibbs_scaling}), (\ref{eq:keq}), (\ref{eq:two_keq}) and (\ref{eq:arrhenius2}) are commonly stated in textbooks.  Our intention is to weave a common mathematical thread between them.  We have surveyed the monographs of \cite{slater}, \cite{johnston}, \cite{zeggeren}, \cite{moore}, \cite{eisenberg}, \cite{smith}, \cite{steinfeld}, \cite{atkins}, \cite{klotz}, \cite{devoe} and \cite{glassman} and verified that, while each lists some subset of these formulae, none of them derive and unify all of these formulae in the manner of the present study.  None of these monographs derive the generalised form of the Arrhenius equation that we present in equation (\ref{eq:arrhenius2}) or the generalisation of the \cite{bs99} analytical solutions we present in \S\ref{subsect:cho}.

\section{General Setup}

In general terms, we consider a chemical reaction involving a pair of reactants (X$_1$ and X$_2$), which produces a pair of products (Z$_1$ and Z$_2$),
\begin{equation}
a_1 \mbox{X}_1 + a_2 \mbox{X}_2 \leftrightarrows b_1 \mbox{Z}_1 + b_2 \mbox{Z}_2,
\label{eq:chemical_reaction_general}
\end{equation}
where $a_1$, $a_2$, $b_1$ and $b_2$ are the stoichiometric coefficients.  The reactants and products may be atoms or molecules of arbitrary stoichiometry.  The forward and reverse reactions are described by the rate coefficients $k_{\rm f}$ and $k_{\rm r}$, respectively.  Whenever we discuss something in general terms, it will always be with reference to the preceding chemical reaction.

\section{Equilibrium Chemistry: Gibbs Free Energy, Equilibrium Constant(s), van't Hoff's Equation and Arrhenius's Equation}

\subsection{The Gibbs free energy}

If we denote the specific internal energy by $U$, the temperature by $T$ and the specific entropy by $S$, then a reasonable guess for the excess energy associated with a chemical reaction is the Helmholtz free energy \citep{slater,moore,eisenberg,atkins,swendsen,devoe},
\begin{equation}
F = U - TS.
\end{equation}
It turns out that this quantity is not general enough because it does not consider the work done on the system.  The general quantity is known as the Gibbs free energy \citep{slater,zeggeren,moore,eisenberg,atkins,swendsen,devoe},
\begin{equation}
G = F + PV,
\end{equation}
where $P$ is the pressure, $V = 1/\rho$ is the specific volume and $\rho$ is the mass density.  

The Gibbs free energy plays a role analogous to the Lagrangian of classical mechanics, which is the difference between the kinetic and potential energies of a system.  Instead of solving Newton's equation directly, one may minimise the Lagrangian, a technique known as the principle of least action.  Gibbs free energy minimisation and chemical kinetics are the chemical analogues to these two techniques. 

In a chemically-active system, the number of particles of each species is generally not a conserved quantity.  If we denote the number of particles associated with the $j$-th species of the system by $N_j$, then the first law of thermodynamics needs to be modified \citep{zeggeren,jacobson,swendsen,glassman},
\begin{equation}
T dS = dU + P ~dV - \sum_j C_j dN_j,
\end{equation}
where $C_j$ is the chemical potential associated with each species.  The sum is performed over all of the species in the system.

By using the definition of $G$ and the product rule, one may show that \citep{smith,klotz,devoe}
\begin{equation}
dG = V dP - S dT + \sum_j C_j dN_j.
\label{eq:gibbs_differential}
\end{equation}
We will now proceed to show that a variety of useful quantities originate from this equation, which is essentially still the first law of thermodynamics.

\subsection{Gibbs free energy minimisation}

Generally, the entropy of a system increases according to the second law of thermodynamics; at constant temperature and pressure, its Gibbs free energy generally decreases and seeks a minimum.  If one is interested in solving for the chemical equilibrium of a network of reactions, then one needs to minimise the Gibbs free energy of the system \citep{zeggeren}.  Within the context of our formalism, we will now elucidate the exact expressions involved in this minimisation.  Generally, we have $C_j = C_j (T,P)$ and equation (\ref{eq:gibbs_differential}) cannot be straightforwardly integrated.  However, at a constant temperature and pressure---which is the typical circumstance under which one performs Gibbs free energy minimisation---equation (\ref{eq:gibbs_differential}) reduces to
\begin{equation}
dG = \sum_j C_j dN_j.
\end{equation}
The integration can be performed trivially to yield \citep{zeggeren,eisenberg,smith,glassman}
\begin{equation}
G = \sum_j C_j N_j.
\label{eq:gibbs_total}
\end{equation}
It is not uncommon to see $G$ being \textit{defined} as the product of the chemical potential and the number of particles of a given species.  Strictly speaking, it is not a definition---rather, it is the expression for $G$ in the isothermal and isobaric limit.  

Equation (\ref{eq:gibbs_total}) is the quantity we need to minimise, but we need additional equations to close the system.  In the absence of nuclear reactions, this arises naturally from the notion that the elemental building blocks of molecules cannot be created or destroyed.  Thus, the number of carbon, hydrogen, oxygen, etc, atoms in a system is invariant between the reactants and the products, whether they exist in their atomic form or are sequestered in molecules.  Mathematically, this set of book-keeping equations takes the form \citep{smith},
\begin{equation}
\sum_j A_{ij} N_j = N^\prime_i.
\label{eq:stoichio_conserved}
\end{equation}
The matrix $A_{ij}$ states the number of atoms of species $i$ present in the molecular species $j$.  The number of atoms of species $i$ is denoted by $N^\prime_i$.

\subsection{The equilibrium constant: more than one definition}

A persistent source of confusion exists in the literature regarding the definition of the equilibrium constant.  Several references list it as being composed of a series of partial pressures associated with the reactants and products and equates it to an exponential term involving the Gibbs free energy (e.g., \citealt{bs99,jacobson,vm11,koppa12,ly13}),
\begin{equation}
K_{\rm eq, literature} = \frac{P^{b_1}_{\rm Z_1} P^{b_2}_{\rm Z_2}}{P^{a_1}_{\rm X_1} P^{a_2}_{\rm X_2}}.
\label{eq:keq_lit}
\end{equation}
Taken at face value, the partial pressure has physical units.  Since the stoichiometric coefficients of the reactants ($a_1+a_2$) and the products ($b_1+b_2$) are generally unequal, $K_{\rm eq, literature}$ must generally have physical units and cannot be equated to an exponential term (which is by definition dimensionless), unless the partial pressures have somehow been normalised.  It is not always explicitly explained that this normalisation has been performed.  Several monographs have previously mentioned this normalisation procedure \citep{fermi,moore,smith,atkins}, but we will now provide a derivation that is consistent with the rest of our formalism.

To derive the equilibrium constant, we return to equation (\ref{eq:gibbs_differential}) and consider it in the limit of $dT=0$ and $dN_j=0$.  If we invoke the ideal gas law ($P = \rho {\cal R} T$), then we obtain (e.g., \citealt{eisenberg,devoe})
\begin{equation}
G = G_0 + {\cal R} T \ln{\left(\frac{P}{P_0}\right)},
\label{eq:gibbs_scaling}
\end{equation}
where $P_0$ is a reference pressure, ${\cal R}$ is the specific gas constant and $G_0 \equiv G(P_0,T)$.  The preceding equation is useful for scaling the Gibbs free energy to other pressures given its value at a reference pressure---it is exactly the equation one has to use when extracting $G$ for different values of $P$ from thermodynamic databases, which typically tabulate values of $G_0$.

We may use the preceding expression to combine the Gibbs free energy of the reactants and products, weighted by their stoichiometric coefficients \citep{slater,eisenberg,smith,klotz,glassman},
\begin{equation}
\Delta G_0 - \Delta G = - {\cal R} T \ln{K_{\rm eq}},
\label{eq:gibbs_keq}
\end{equation}
where we have defined
\begin{equation}
\begin{split}
\Delta G &\equiv b_1 G_{\rm Z_1} + b_2 G_{\rm Z_2} - a_1 G_{\rm X_1} - a_2 G_{\rm X_2}, \\
\Delta G_0 &\equiv b_1 G_{\rm Z_1,0} + b_2 G_{\rm Z_2,0} - a_1 G_{\rm X_1,0} - a_2 G_{\rm X_2,0}. \\
\end{split}
\end{equation}
What is interesting is that this first-principles approach naturally yields the definition for the equilibrium constant (e.g., \citealt{klotz}),
\begin{equation}
K_{\rm eq} \equiv \frac{ \left( P_{\rm Z_1}/P_0 \right)^{b_1} \left( P_{\rm Z_2}/P_0 \right)^{b_2} }{ \left( P_{\rm X_1}/P_0 \right)^{a_1} \left( P_{\rm X_2}/P_0 \right)^{a_2} }.
\label{eq:keq}
\end{equation}
Notice that this equilibrium constant is naturally dimensionless; its derivation is similar to the ones given in \cite{moore}, \cite{smith}, \cite{atkins} and \cite{devoe}, who obtained it in terms of chemical potentials and activities.  The factors of $P_0$ appear without being inserted in an ad hoc manner.

Physically, the system adjusts itself until it reaches chemical equilibrium, which occurs when $\Delta G = 0$.  Let the reference state, characterised by $P_0$, \textit{not} be in equilibrium, such that $\Delta G_0 \ne 0$.  If one is referring to a molecule, then $\Delta G_0$ is the energy needed to construct it from its constituent atoms---it is the Gibbs free energy of formation.  If one is referring to mixtures of molecules, then $\Delta G_0$ is the difference in the Gibbs free energies of formation between the reactants and products.  Equation (\ref{eq:gibbs_keq}) naturally yields the relationship between $\Delta G_0$ and $K_{\rm eq}$ \citep{slater,johnston,zeggeren,moore,eisenberg,smith,steinfeld,atkins,klotz,devoe,glassman},
\begin{equation}
K_{\rm eq} = \exp{\left(-\frac{\Delta G_0}{{\cal R} T} \right)}.
\end{equation}

For example, while equations (A1) and (A6) of \cite{bs99}, equation (7) of \cite{vm11}, equation (5) of \cite{koppa12} and equation (2) of \cite{ly13} do not explicitly mention the factors of $P_0$ needed to render $K_{\rm eq}$ dimensionless, it is common practice to omit these reference-pressure terms in standard treatments of chemical equilibria.

We now seek another possible definition of the equilibrium constant.  Let the number density be generally represented by $n$; self-explanatory subscripts relate it to the appropriate reactant or product.  In chemical equilibrium, we expect the forward and reverse rate coefficients to be related as follows \citep{moore},
\begin{equation}
k_{\rm f} n_{\rm X_1}^{a_1} n_{\rm X_2}^{a_2} = k_{\rm r} n_{\rm Z_1}^{b_1} n_{\rm Z_2}^{b_2}.
\end{equation}
A plausible, alternative definition for the equilibrium constant is \citep{johnston,moore,steinfeld}
\begin{equation}
K_{\rm eq}^\prime \equiv \frac{k_{\rm f}}{k_{\rm r}}.
\end{equation}
Note that since $k_{\rm f}$ and $k_{\rm r}$ generally do not possess the same physical units, $K_{\rm eq}^\prime$ is expected to be dimensional.

We may relate our two definitions of the equilibrium constant \citep{atkins},
\begin{equation}
K_{\rm eq}^\prime = K_{\rm eq} \left( k_{\rm B} T \right)^{a_1+a_2-b_1-b_2} P_0^{b_1+b_2-a_1-a_2},
\label{eq:two_keq}
\end{equation}
with $k_{\rm B}$ being the Boltzmann constant.  As has been pointed out by \cite{vm11}, the ``pressure correction term" (which is really a temperature correction term) is sometimes missed by other workers.  It is less well-known that this correction term has already been elucidated by \cite{fermi}.  It vanishes when $a_1+a_2=b_1+b_2$ and we have $K_{\rm eq} = K^\prime_{\rm eq}$ (which is commonly, but not always, true).  

In other words, $K^\prime_{\rm eq}$ is used to ``reverse" the forward rate coefficients, but it is $K_{\rm eq}$ that relates it to the Gibbs free energy.  Equation (\ref{eq:two_keq}) relates them properly.  In the literature, what we have defined as $K_{\rm eq}$ and $K^\prime_{\rm eq}$ are often denoted, respectively, by $K_P$ and $K_{\rm eq}$ instead (e.g., \citealt{vm11}), although such an approach is not universally adopted (e.g., \citealt{ly13}).

\subsection{The van't Hoff equation}

If we differentiate equation (\ref{eq:gibbs_keq}) with respect to the temperature, we obtain the van't Hoff equation \citep{smith,jacobson},
\begin{equation}
\frac{\partial \left( \ln{K_{\rm eq}} \right)}{\partial T} = \frac{\Delta G_0}{{\cal R} T^2}.
\label{eq:vant_hoff}
\end{equation}
Note that we are allowed to go from equation (\ref{eq:gibbs_keq}) to (\ref{eq:vant_hoff}) only because we have \textit{constructed} $\Delta G_0$ to be isothermal.  

In most incarnations of the van't Hoff equation, it is the change in enthalpy, rather than the Gibbs free energy, which is stated \citep{moore,smith,jacobson,devoe}.  If the system is isothermic and adiabatic, then these two statements are equivalent \citep{slater}.

\subsection{The Arrhenius equation: rate coefficients and activation energies}

\begin{figure}
\begin{center}
\includegraphics[width=\columnwidth]{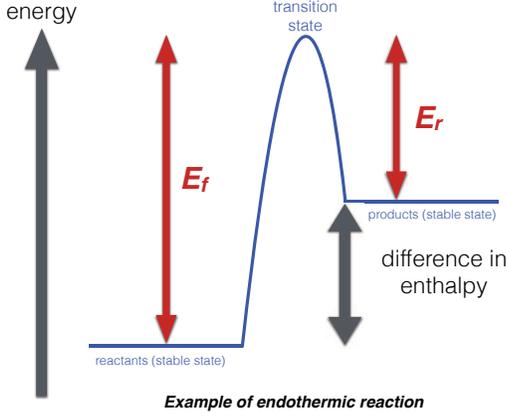}
\end{center}
\caption{Schematic depicting the relationship between the activation energies and the change in enthalpy.}
\vspace{0.1in}
\label{fig:activation}
\end{figure}

We next derive the expressions for the rate coefficients.  At this point, we need to invoke the notion of the activation energy, which is the energy barrier associated with a forward or reverse reaction.  One may think of the reactants and products as being two different stable states residing at different energy levels.  To transition from one state to the other requires that one surmounts an energy barrier, which is the activation energy (Figure \ref{fig:activation}).  The barrier of the activation energy originates from the need to overcome bond strengths and the requirement that the reactants have specific orientations during a collision.  For a single reaction, the difference between the activation energies of the forward and reverse reactions is the change in the enthalpy,
\begin{equation}
\Delta G_0 = E_{\rm f} - E_{\rm r} - T \Delta S_0,
\end{equation}
where $E_{\rm f}$ and $E_{\rm r}$ are the activation energies associated with the forward and reverse reactions, respectively, and $\Delta S_0$ is the change in entropy at the reference pressure.  The preceding expression allows us to cast the adjectives ``exothermic" and ``endothermic" in more precise, mathematical terms.  If the activation energy of the forward reaction exceeds that of the reverse one, then one needs to inject energy into the system for it to proceed, i.e., $E_{\rm f} - E_{\rm r} > 0$.  One refers to this as an endothermic reaction.  Reactions with $E_{\rm f} - E_{\rm r} < 0$ are exothermic.

By combining the expressions for $K_{\rm eq}$ and $K^\prime_{\rm eq}$, we obtain
\begin{equation}
\begin{split}
\ln{k_{\rm f}} - \ln{k_{\rm r}} =& - \frac{E_{\rm f} - E_{\rm r}}{{\cal R} T} + \frac{\Delta S_0}{{\cal R}} \\
&+ \left( a_1+a_2-b_1-b_2 \right) ~\ln{\left( \frac{k_{\rm B} T}{P_0} \right)}.
\end{split}
\end{equation}
The symmetries inherent in the preceding equation suggests that it may have been constructed from two independent governing equations for the rate coefficients \citep{upa}.  Mathematically, ``splitting" this equation is a degenerate endeavor and is not rigorous.  To persist in this endeavor, we have to appeal to physics.  First, we expect that the governing equations for $k_{\rm f}$ and $k_{\rm r}$ must enjoy a large degree of symmetry between them.  Second, we expect $k_{\rm f}$ and $k_{\rm r}$ to be associated with $E_{\rm f}$ and $E_{\rm r}$, respectively.  Thus, a plausible guess is that the preceding equation originated from the difference between these two equations,
\begin{equation}
\begin{split}
\ln{k_{\rm f}} &= - \frac{E_{\rm f}}{{\cal R} T} + c_{\rm f} \ln{T} + \frac{c_{\rm f}^\prime \Delta S_0}{{\cal R}} + c_{\rm f}^{\prime \prime},  \\
\ln{k_{\rm r}} &= - \frac{E_{\rm r}}{{\cal R} T} + c_{\rm r} \ln{T} + \frac{c_{\rm r}^\prime \Delta S_0}{{\cal R}} + c_{\rm r}^{\prime \prime}. \\
\end{split}
\label{eq:arrhenius}
\end{equation}
The coefficients $c_{\rm f}$ and $c_{\rm r}$ cannot be stated uniquely.  For example, we can have $c_{\rm f} = a_1+ a_2$ and $c_{\rm r} = b_1+ b_2$; we may also have $c_{\rm f} = - b_1 - b_2$ and $c_{\rm r} = -a_1 - a_2$.  This mathematical freedom implies that $c_{\rm f}$ and $c_{\rm r}$ may take on a range of values and may be positive or negative.

Finally, we end up with the Arrhenius equations, 
\begin{equation}
\begin{split}
k_{\rm f} &= A_{\rm f} ~T^{c_{\rm f}} ~\exp{\left( - \frac{E_{\rm f}}{{\cal R}T} \right)}, \\
k_{\rm r} &= A_{\rm r} ~T^{c_{\rm r}} ~\exp{\left( - \frac{E_{\rm r}}{{\cal R}T} \right)}, \\
\end{split}
\label{eq:arrhenius2}
\end{equation}
where we necessarily have
\begin{equation}
\begin{split}
&c_{\rm f} - c_{\rm r} = a_1+a_2-b_1-b_2, \\
&c^\prime_{\rm f} - c^\prime_{\rm r} = 1, \\
&c_{\rm f}^{\prime \prime} - c_{\rm r}^{\prime \prime} = \left( a_1+a_2-b_1-b_2 \right) ~\ln{\left( \frac{k_{\rm B}}{P_0} \right)}.
\end{split}
\end{equation}
The pre-exponential factor $A_{\rm f}$ absorbs terms associated with $c^\prime_{\rm f}$, $c_{\rm f}^{\prime \prime}$ and $\Delta S_0$; its counterpart, $A_{\rm r}$, does the same for $c^\prime_{\rm r}$, $c_{\rm r}^{\prime \prime}$ and $\Delta S_0$.  Absorbing the entropy into the pre-exponential factors was previously noted by \cite{yung}, but our derivation is more general as it involves $c^\prime_{\rm f}$ and $c^\prime_{\rm r}$.  We note that one may also use the van't Hoff equation as a starting point for the derivation.

Traditionally, derivations or statements of the Arrhenius equation include only the exponential term involving the activation energy \citep{johnston,moore,steinfeld,yung,jacobson,atkins}.  They omit the power-law terms and tag them on, after the fact (e.g., \citealt{jacobson}), partially as a means of using them as fitting functions for experimental data.  Our derivation demonstrates that there is a sound basis to including these terms.  Thus, the Arrhenius equations attain a status that is elevated above that of mere ad hoc fitting functions.  Typically, $c_{\rm f} = 0$ and $c_{\rm r}=0$ suffice for low temperatures; ``non-Arrhenius" behavior, where $c_{\rm f} \ne 0$ and $c_{\rm r} \ne 0$, is important at high temperatures \citep{glassman}.

Kinetic theory states that the rate coefficient is given by $\langle \sigma_{\rm coll} v_{\rm rel} \rangle$, where $\sigma_{\rm coll}$ is the cross section for collisions between the reactants and $v_{\rm rel}$ is the relative velocity between them.  If $\sigma_{\rm coll}$ is independent of the relative velocity, then $\langle \sigma_{\rm coll} v_{\rm rel} \rangle \propto T^{1/2}$ \citep{glassman}.  If $\sigma_{\rm coll}$ depends on the relative velocity, then more general power-law dependences on $T$ are possible.

The Arrhenius equations do not account for three-body reactions.  When the number density of the third body is low, the reaction rate is linearly proportional to it.  As it increases, a point is reached where the reaction rate saturates to a limiting value.  Fitting functions for implementing this saturation effect have previously been given by, for example, \cite{vm11}.

\section{Analytical Models of Atmospheric Chemistry}

The formalism and concepts we have established may be highlighted via a set of analytical models.

\subsection{Pure hydrogen}
\label{subsect:hydrogen}

\begin{figure}
\begin{center}
\includegraphics[width=\columnwidth]{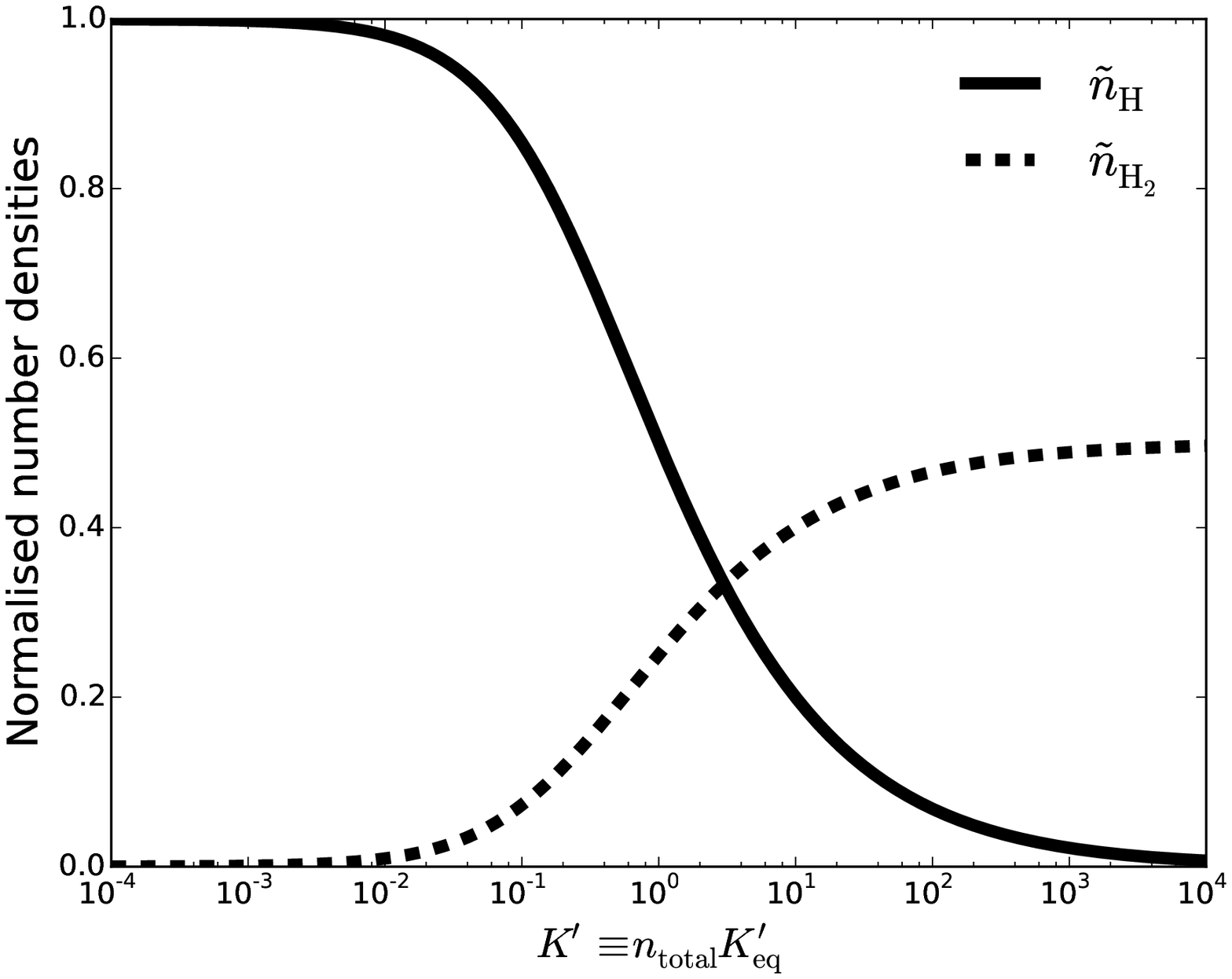}
\includegraphics[width=\columnwidth]{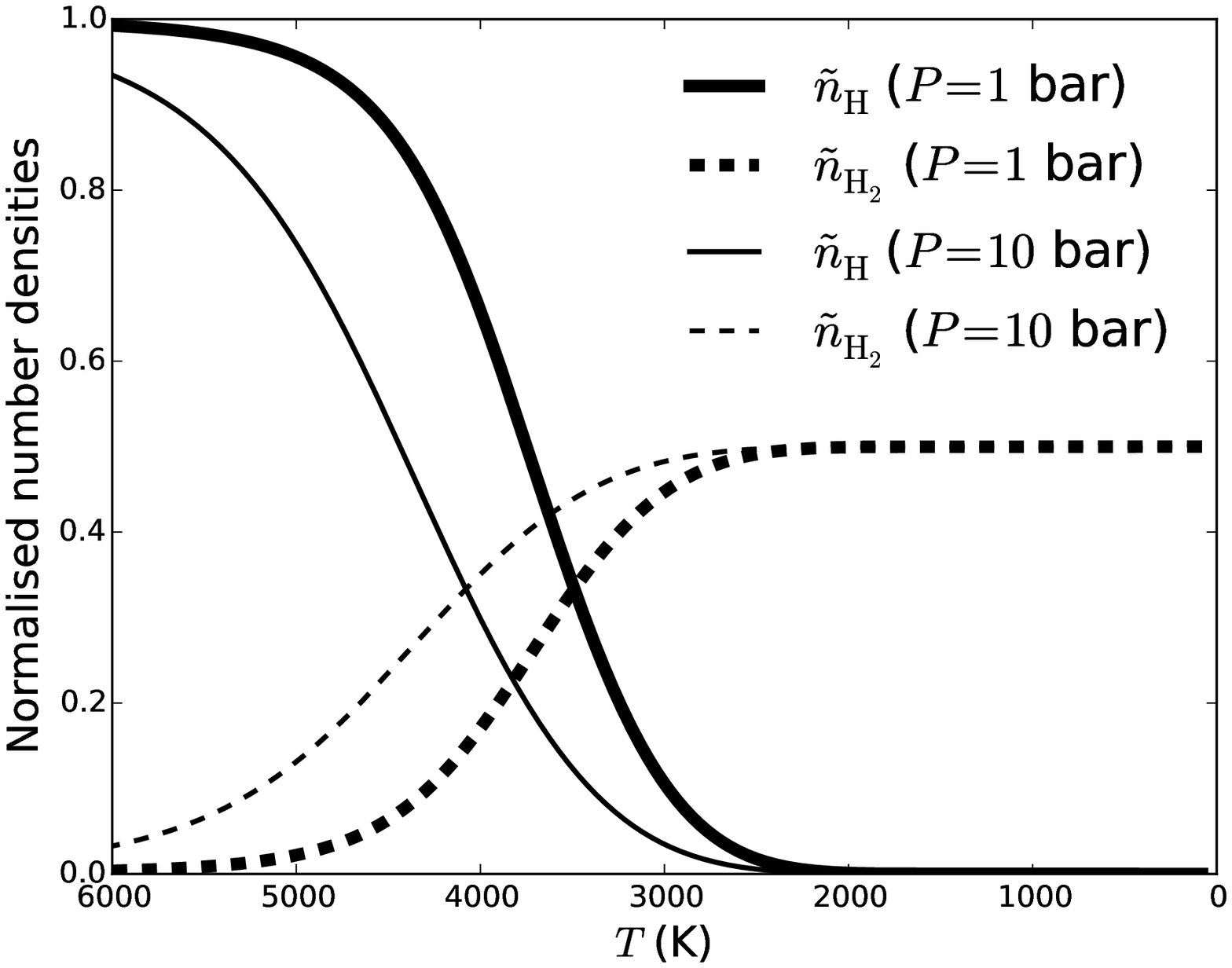}
\end{center}
\caption{Abundances of atomic and molecular hydrogen, normalised by the total number density, as a function of the normalised equilibrium constant (top panel) and temperature (bottom panel).  In the top panel, $K^\prime$ is a proxy for the temperature; larger $K^\prime$ values correspond to lower temperatures.}
\vspace{0.1in}
\label{fig:hydrogen}
\end{figure}

For completeness and as the simplest example, we consider a system consisting purely of hydrogen in its atomic and molecular forms,
\begin{equation}
2\mbox{H} + \mbox{M} \leftrightarrows \mbox{H}_2 + \mbox{M},
\label{eq:hydrogen_reaction}
\end{equation}
where M is a third body of arbitrary stoichiometry.

Using our formalism for the evolution equations (see Appendix \ref{append:kinetics}), we may write down
\begin{equation}
\begin{split}
\frac{1}{2} \frac{\partial n_{\rm H}}{\partial t} &= -n_{\rm H}^2 n_{\rm M} k_{\rm f} + n_{\rm H_2} n_{\rm M} k_{\rm r}, \\
\frac{\partial n_{\rm H_2}}{\partial t} &= n_{\rm H}^2 n_{\rm M} k_{\rm f} - n_{\rm H_2} n_{\rm M} k_{\rm r}. \\
\end{split}
\end{equation}
If we add these equations and perform the integration, we end up with
\begin{equation}
n_{\rm H} + 2 n_{\rm H_2} = n_{\rm total}.
\label{eq:pure_hydrogen}
\end{equation}
This is already a demonstration that the correction factor (the reciprocal of the stoichiometric coefficient) is essential \citep{johnston,steinfeld}, if one desires to get the book-keeping between the hydrogen atoms and molecules correct.  Here, $n_{\rm total}$ is the total number of particles in the system.

In chemical equilibrium, the (dimensional) equilibrium constant of the reaction is
\begin{equation}
K^\prime_{\rm eq} = \frac{n_{\rm H_2}}{n_{\rm H}^2}.
\end{equation}
If we plug this expression back into equation (\ref{eq:pure_hydrogen}) and define $K^\prime \equiv K^\prime_{\rm eq} n_{\rm total}$, we may solve for the (normalised) number density of atomic hydrogen \citep{gail},
\begin{equation}
\tilde{n}_{\rm H} \equiv \frac{n_{\rm H}}{n_{\rm total}} = \frac{-1 + \left( 1 + 8 K^\prime \right)^{1/2}}{4 K^\prime}.
\end{equation}
The preceding expression is similar, but not identical, to that presented in \cite{barshay78}.

Since $K^\prime \propto K^\prime_{\rm eq}P/T$, one may argue that increasing $K^\prime$ values correspond to decreasing temperatures.  Figure \ref{fig:hydrogen} shows the curves of $\tilde{n}_{\rm H}$ and $\tilde{n}_{\rm H_2} \equiv n_{\rm H_2}/n_{\rm total}$.  As expected, molecular hydrogen prevails at low temperatures.  So far, our toy model does not allow us to define what ``low" is, as we have not related $K^\prime$ to $T$ and $P$.  Appendix \ref{append:gibbs} lists the Gibbs free energies used to make this conversion.  In Figure \ref{fig:hydrogen}, we include a separate set of calculations where $\tilde{n}_{\rm H}$ and $\tilde{n}_{\rm H_2}$ are shown as functions of temperature and pressure.  At $T \lesssim 3000$ K, hydrogen exists predominantly in its molecular form.

\subsection{Carbon, hydrogen and oxygen: methane, water, carbon monoxide and acetylene}
\label{subsect:cho}

Inspired by the work of \cite{bs99}, we seek to generalise our toy model of a system with pure hydrogen to one that contains carbon (C), hydrogen (H) and oxygen (O), albeit only in gaseous form, and any carbon-to-oxygen ratio (C/O).  We wish to compute the relative abundances of the resulting molecules: methane (CH$_4$), water (H$_2$O), carbon monoxide (CO) and acetylene (C$_2$H$_2$).

We consider the reaction of methane with water to form carbon monoxide and molecular hydrogen \citep{bs99,lodders02,moses11},
\begin{equation}
\mbox{CH}_4 + \mbox{H}_2\mbox{O} \leftrightarrows \mbox{CO} + 3 \mbox{H}_2.
\label{eq:chemical_reaction_example}
\end{equation}
The formulae presented in the appendix of \cite{bs99} consider only this reaction and thus are unable to represent carbon-rich atmospheres, where a variety of hydrocarbons are present at high temperatures \citep{lodders02,madhu12,venot15}.  If these hydrocarbons are excluded, then one gets the spurious result that methane is always the dominant carbon carrier at high temperatures and in carbon-rich situations.  Our desire for an analytical model does not allow us to include all of the hydrocarbons that are expected to form.  Instead, we assume that acetylene is the dominant hydrocarbon and include it via the following reaction \citep{lodders02,moses11},
\begin{equation}
2\mbox{CH}_4 \leftrightarrows \mbox{C}_2\mbox{H}_2 + 3 \mbox{H}_2.
\label{eq:chemical_reaction_example_2}
\end{equation}
If nitrogen is present, we expect hydrogen cyanide (HCN) to form as well \citep{madhu12}, but in the interest of algebraic tractability we will not include it.  Furthermore, \cite{venot15} have shown using calculations of chemical kinetics that acetylene and hydrogen cyanide are the dominant hydrocarbons in carbon-rich atmospheres.

In reality, both reactions are net reactions that consist of large networks of individual reactions, some of which produce transient species en route to the products.  We assume that hydrogen exists mostly in its molecular form, such that the partial pressure of H$_2$ is, to a good approximation, the total pressure ($P$) of the system.  Atomic hydrogen is expected to introduce only a small correction to $P = n_{\rm H_2} k_{\rm B} T$.  This simplification essentially removes the need for an additional equilibrium constant to account for the atomic to molecular transition (and vice versa) of hydrogen, as was described in Section \ref{subsect:hydrogen}.

\begin{figure}
\begin{center}
\includegraphics[width=\columnwidth]{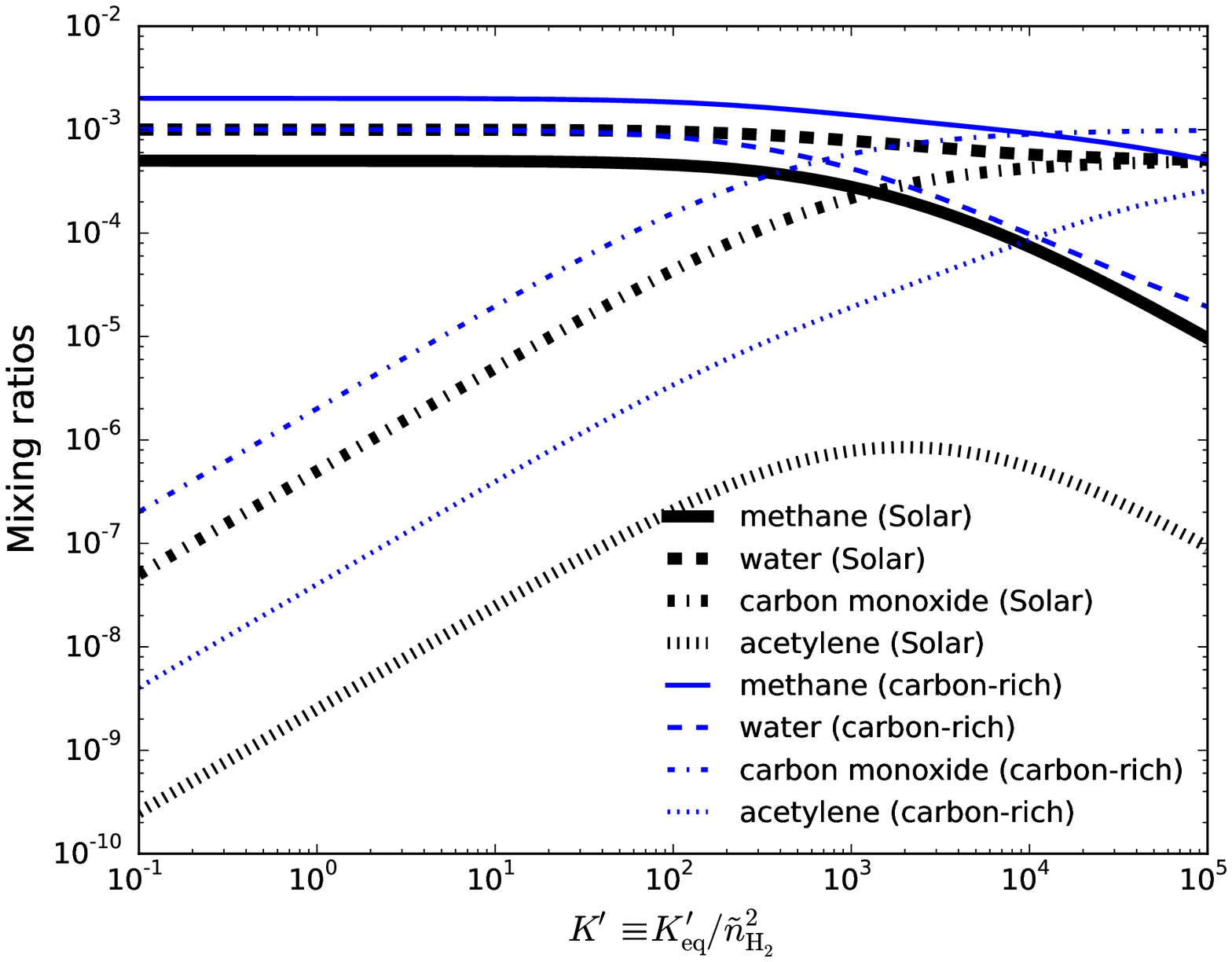}
\includegraphics[width=\columnwidth]{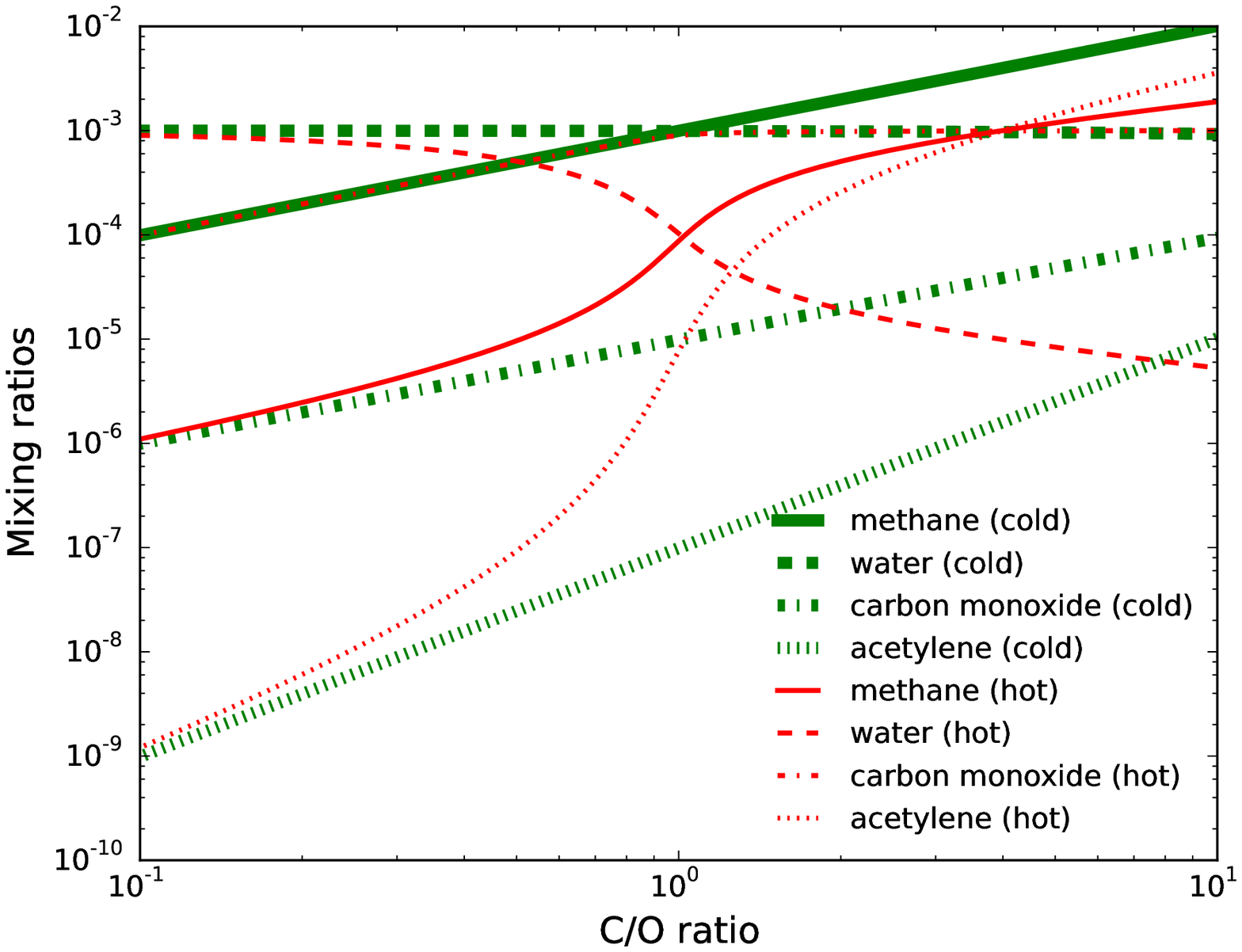}
\end{center}
\caption{Mixing ratios of methane, water, carbon monoxide and acetylene.  Again, $K^\prime$ is a proxy for the temperature, but larger $K^\prime$ values correspond to higher temperatures.  We have set $\tilde{n}_{\rm O} = 5 \times 10^{-4}$ as this is the approximate value of the Sun's photospheric oxygen abundance \citep{lodders03}.  Top panel: mixing ratios as a function of $K^\prime$.  The ``Solar" and ``carbon-rich" cases correspond to $\tilde{n}_{\rm C}/\tilde{n}_{\rm O} = 0.5$ and $\tilde{n}_{\rm C}/\tilde{n}_{\rm O} = 2$, respectively.  Bottom panel: mixing ratios as a function of the carbon-to-oxygen ratio ($\tilde{n}_{\rm C}/\tilde{n}_{\rm O}$).  The ``cold" and ``hot" cases correspond to $K^\prime = 10$ and $K^\prime = 10^5$, respectively.  For illustration, we have set $K^\prime_2/K^\prime = 10^{-2}$.}
\vspace{0.1in}
\label{fig:CHO}
\end{figure}

\begin{figure}
\begin{center}
\includegraphics[width=\columnwidth]{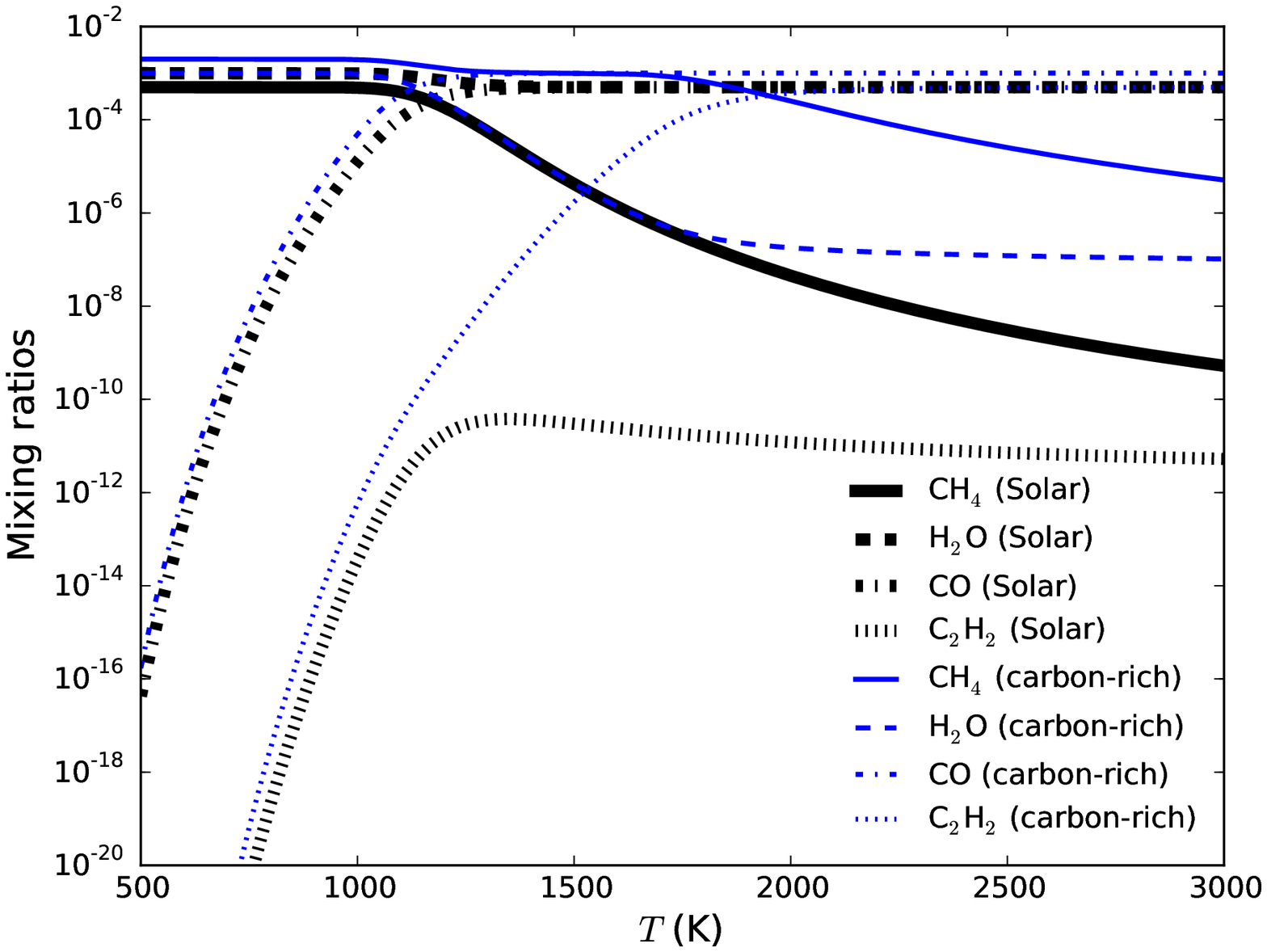}
\includegraphics[width=\columnwidth]{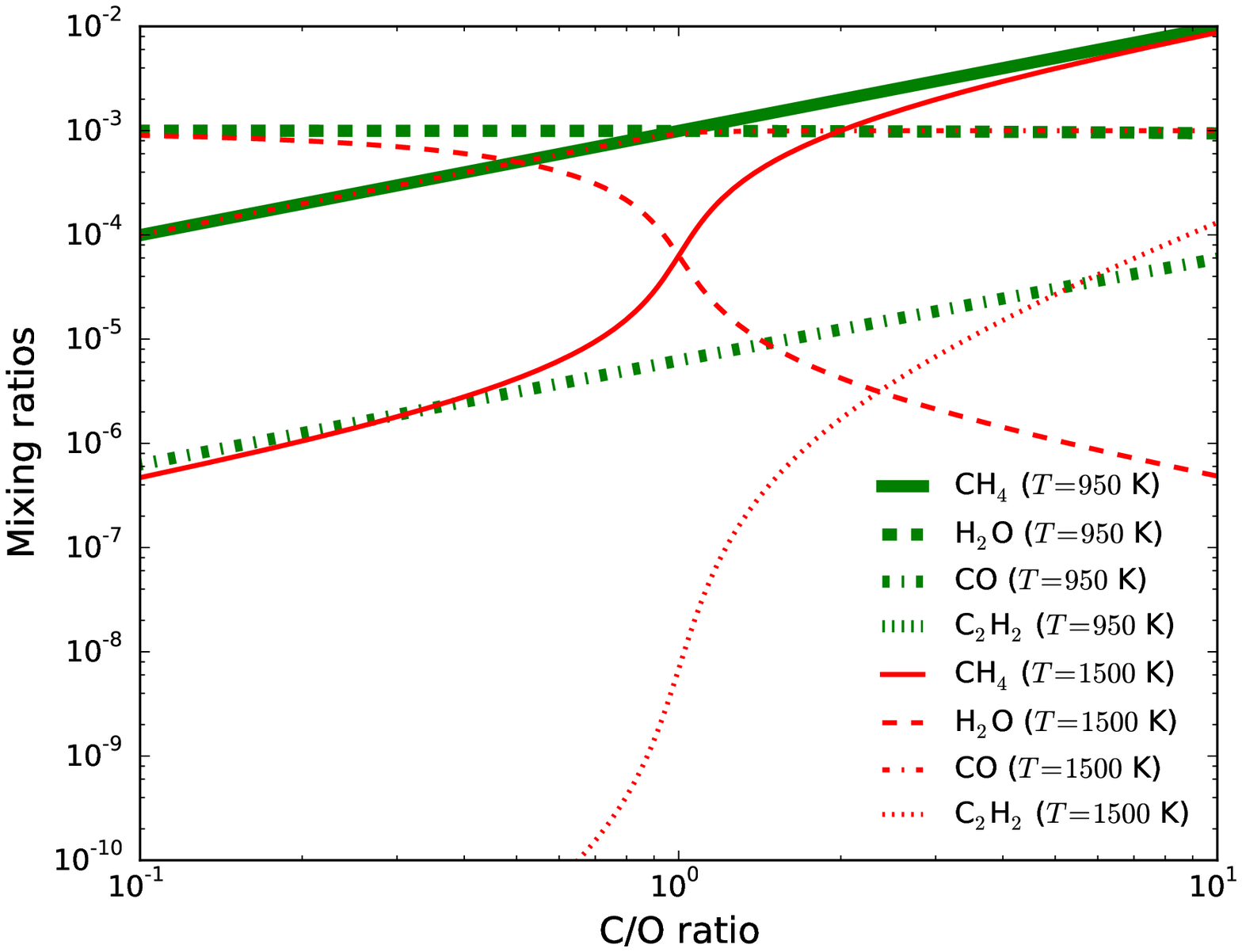}
\end{center}
\caption{Same as Figure \ref{fig:CHO}, but with the equilibrium constants being related to temperature and pressure via the Gibbs free energies taken from the JANAF database.  For illustration, we set $P=1$ bar.  Here, the ``cold" and ``hot" cases correspond to $T=950$ K and $T=1500$ K, respectively.  For the bottom panel, note that acetylene has a mixing ratio below $10^{-10}$ (the lower limit of the vertical axis) for the cold case.}
\vspace{0.1in}
\label{fig:CHO_2}
\end{figure}

The dimensional equilibrium constant of the reaction described in equation (\ref{eq:chemical_reaction_example}) is
\begin{equation}
K^\prime_{\rm eq} = \frac{n_{\rm CO} n_{\rm H_2}^3}{n_{\rm CH_4} n_{\rm H_2O}} = \frac{\tilde{n}_{\rm CO} n_{\rm H_2}^2}{\tilde{n}_{\rm CH_4} \tilde{n}_{\rm H_2O}},
\end{equation}
while that of the reaction in equation (\ref{eq:chemical_reaction_example_2}) is
\begin{equation}
K^\prime_{\rm eq,2} = \frac{n_{\rm C_2H_2} n_{\rm H_2}^3}{n_{\rm CH_4}^2} = \frac{\tilde{n}_{\rm C_2H_2} n_{\rm H_2}^2}{\tilde{n}_{\rm CH_4}^2}.
\end{equation}
Analogous to the case study of pure hydrogen, we have defined
\begin{equation}
K^\prime \equiv \frac{K^\prime_{\rm eq}}{n_{\rm H_2}^2}, ~K^\prime_2 \equiv \frac{K^\prime_{\rm eq,2}}{n_{\rm H_2}^2},
\end{equation}
but we note that since $K^\prime \propto K^\prime_{\rm eq} T^2 / P^2$, we expect $K^\prime$ to increase with temperature, opposite from the trend associated with the pure-hydrogen system.  We will again use $K^\prime$ as a proxy for the temperature.  An important limitation of our model is the difficulty with relating $K^\prime$ and $K^\prime_2$, because this requires us to explicitly state the functional forms of the change in Gibbs free energies of the two reactions.  We first make the simplest assumption: that $K^\prime_2/K^\prime$ is a constant; we will discuss the implications of this assumption later.  While we could certainly specify the temperature dependence of $K^\prime$ and $K^\prime_2$---which is what \cite{bs99} did---we initially choose not to so as to understand what such a simple model would teach us.  We will see shortly that the simplicity yields an important insight, which is that we recover most of the qualitative trends simply by using the equilibrium constants as proxies for the temperature.

The conservation of particles, as described in equation (\ref{eq:stoichio_conserved}), states that
\begin{equation}
\begin{split}
&n_{\rm CH_4} + n_{\rm CO} + 2 n_{\rm C_2H_2} = n_{\rm C}, \\
&n_{\rm H_2O} + n_{\rm CO} = n_{\rm O}, \\
&4n_{\rm CH_4} + 2 n_{\rm H_2O} + 2 n_{\rm C_2H_2} + 2n_{\rm H_2} = n_{\rm H}.
\end{split}
\end{equation}
These equations may be manipulated to obtain
\begin{equation}
\begin{split}
&\tilde{n}_{\rm CH_4} + \tilde{n}_{\rm CO} + 2 \tilde{n}_{\rm C_2H_2} \\
&= \tilde{n}_{\rm C} \left( 4\tilde{n}_{\rm CH_4} + 2 \tilde{n}_{\rm H_2O} + 2 \tilde{n}_{\rm C_2H_2} + 2 \right), \\
&\tilde{n}_{\rm H_2O} + \tilde{n}_{\rm CO} = \tilde{n}_{\rm O} \left( 4\tilde{n}_{\rm CH_4} + 2 \tilde{n}_{\rm H_2O} + 2 \tilde{n}_{\rm C_2H_2} + 2 \right).
\end{split}
\end{equation}
Note that the number densities of the molecules marked by tildes have been normalised by $n_{\rm H_2}$, while those of the atoms have been normalised by $n_{\rm H}$.  The former are the mixing ratios, while the latter are the normalised elemental abundances.

With two particle conservation equations and the expressions for $K^\prime$ and $K^\prime_2$, we have four equations and four unknowns.  They can be manipulated to yield a cubic equation for the mixing ratio of methane,
\begin{equation}
{\cal C}_0 \tilde{n}_{\rm CH_4}^3 + {\cal C}_1 \tilde{n}_{\rm CH_4}^2 + {\cal C}_2 \tilde{n}_{\rm CH_4} + {\cal C}_3 = 0,
\end{equation}
which has the coefficients,
\begin{equation}
\begin{split}
{\cal C}_0 =& 2K^\prime K^\prime_2 \left( \tilde{n}_{\rm O} - \tilde{n}_{\rm C} + 1 \right),\\
{\cal C}_1 =& K^\prime \left( 4 \tilde{n}_{\rm O} - 4 \tilde{n}_{\rm C} + 1 \right) \\
&- K^\prime_2 \left[ 4 \tilde{n}_{\rm O} \tilde{n}_{\rm C} + 2 \left( 1 - 2 \tilde{n}_{\rm O} \right) \left( \tilde{n}_{\rm C} - 1 \right) \right],\\
{\cal C}_2 =& 2K^\prime \left( \tilde{n}_{\rm O} - \tilde{n}_{\rm C} \right) - 4\tilde{n}_{\rm C} - 2\tilde{n}_{\rm O} + 1,\\
{\cal C}_3 =& - 2\tilde{n}_{\rm C}.
\end{split}
\end{equation}
While analytical solutions do exist for cubic equations, they possess multiple branches---some of which are complex---depending on tedious combinations of the values of ${\cal C}_0$, ${\cal C}_1$, ${\cal C}_2$ and ${\cal C}_3$ \citep{nr}.  Since it is difficult to determine a prior which solution branch $\tilde{n}_{\rm CH_4}$ is described by, we elect to solve the cubic equation using standard, canned numerical routines for solving polynomial equations.  The other mixing ratios can be obtained via
\begin{equation}
\begin{split}
\tilde{n}_{\rm H_2O} &= \frac{2 \tilde{n}_{\rm O} \left( K^\prime_2 \tilde{n}_{\rm CH_4}^2 + 2\tilde{n}_{\rm CH_4} + 1 \right)}{1 + K^\prime \tilde{n}_{\rm CH_4} - 2\tilde{n}_{\rm O}}, \\
\tilde{n}_{\rm CO} &= K^\prime \tilde{n}_{\rm CH_4} \tilde{n}_{\rm H_2O}, \\
\tilde{n}_{\rm C_2H_2} &= K^\prime_2 \tilde{n}_{\rm CH_4}^2.
\end{split}
\end{equation}

For completeness, we note that when acetylene is absent ($K^\prime_2=0$), the solution can be easily written down,
\begin{equation}
\tilde{n}_{\rm CH_4} = \frac{- {\cal C}_2 + \left( {\cal C}_2^2 - 4 {\cal C}_1 {\cal C}_3 \right)^{1/2}}{2 {\cal C}_1}.
\end{equation}
Notice how the coefficient ${\cal C}_0$, and thus $K^\prime_2$, controls the extent to which the mixing ratio of methane is described by a quadratic versus cubic equation.  Physically, we expect that at low temperatures ($K^\prime \ll 1$), the mixing ratio of acetylene is negligible.  In this limit ($K^\prime, K^\prime_2 \ll 1$), we have $\tilde{n}_{\rm CH_4} \approx 2 \tilde{n}_{\rm C}$ and $\tilde{n}_{\rm H_2O} \approx 2 \tilde{n}_{\rm O}$.  These asymptotic solutions explain the relatively simple behavior of the mixing ratios at low temperatures, as seen in Figure \ref{fig:CHO}.  It also offers an easy explanation for methane and water switching roles as the dominant molecule when the carbon-to-oxygen ratio is exactly unity, as noted by \cite{koppa12}, since $\tilde{n}_{\rm CH_4}/\tilde{n}_{\rm H_2O} \approx \tilde{n}_{\rm C}/\tilde{n}_{\rm O}$.

Figure \ref{fig:CHO} shows the mixing ratios of methane, water, carbon monoxide and acetylene as functions of $K^\prime$.  Our analytical model produces the following, salient trends.
\begin{itemize}

\item When the atmosphere has a solar abundance of elements, water is always more abundant than methane \citep{bs99,moses13a}.  At low temperatures, methane dominates carbon monoxide as the carrier of carbon \citep{prinn77,barshay78}; this trend reverses at high temperatures \citep{bs99,lodders02}.

\item When the atmosphere is carbon-rich, methane is the dominant molecule but has to compete with acetylene in some circumstances \citep{madhu12,moses13a,venot15}.  Water is the dominant oxygen carrier only at low temperatures, superceded by carbon monoxide at high temperatures \citep{madhu12,moses13a,moses13b}.

\item Cold atmospheres are always methane-rich at the expense of carbon monoxide, regardless of the C/O \citep{madhu12}.  The abundance of water is essentially constant across C/O \citep{madhu12}.

\item Hot atmospheres exhibit more complex behavior, in that they are methane-poor and water-rich when $\mbox{C/O}<1$ \citep{madhu12,moses13a}.  For $\mbox{C/O}>1$, they become methane-rich and water-poor \citep{madhu12,moses13a} with methane dominating carbon monoxide as the carrier of carbon when C/O becomes sufficiently larger than unity \citep{madhu12}.  When C/O is large enough, acetylene overtakes methane as the dominant carrier of carbon \citep{madhu12}.

\end{itemize}

These trends are in agreement with the numerical calculations of equilibrium chemistry presented in \cite{madhu12} and \cite{moses13a}, but with one exception.  We have assumed $K^\prime_2/K^\prime=10^{-2}$; higher values would produce the unphysical result that acetylene dominates carbon monoxide, even at low temperatures (not shown).  The cold model in the lower panel of Figure \ref{fig:CHO} shows an overabundance of acetylene, which is in disagreement with Figure 2 of \cite{madhu12}.  This discrepancy arises from the fact that $K^\prime_2/K^\prime$ is not a constant and must possess a (steep) temperature dependence.

To investigate this discrepancy further, we stop treating $K^\prime$ and $K^\prime_2$ as free parameters and instead relate them to temperature and pressure via the Gibbs free energy tabulated in the JANAF database (\texttt{http://kinetics.nist.gov/janaf/}).  Our implementation of this procedure is described in \cite{hl15}.  In Figure \ref{fig:CHO_2}, we recalculate the models in Figure \ref{fig:CHO}.  We see that the basic trends previously discussed are preserved, although the curves display quantitative differences as expected.  The mixing ratios versus C/O match surprisingly well even at a quantitative level.  The previous result regarding acetylene is verified to be an artifact of assuming $K^\prime_2/K^\prime$ to be constant.  We further verified that acetylene becomes dominant over methane only for carbon-rich atmospheres with $T \ge 2000$ K (not shown), in agreement with \cite{madhu12}.

Overall, it is surprising how well our model is able to reproduce the main trends of mixing ratios versus the carbon-to-oxygen ratio.  It is surprising because this rich variety of behavior originates from the chemical analogue of geometry---it is merely stoichiometric book-keeping \citep{lodders02}.  The dependence of the normalised equilibrium constants on temperature is a distraction if all one seeks is to understand these trends in a qualitative sense.

\section{Summary}

We have presented a unified, novel, self-consistent formalism for understanding the atmospheric chemistry of exoplanets from the viewpoint of an astrophysicist.  In doing so, we addressed ambiguities associated with the equilibrium constant and obtained a novel derivation of the Arrhenius equation.  We also generalised previous work on analytical models of systems in chemical equilibrium with carbon, hydrogen and oxygen and showed that they reproduce several key trends published in the literature and computed using more sophisticated numerical calculations.  We anticipate that such models are useful for inclusion in retrieval models of exoplanetary atmospheres to maintain their chemical plausibility as a first approach \citep{benneke15}.


\appendix

\section{Chemical Kinetics}
\label{append:kinetics}

For completeness, we restate the formalism concerning chemical kinetics.

\subsection{Evolution equations}

Unlike Gibbs free energy minimisation, chemical kinetics is the treatment of a network of reactions as a system of mass conservation equations.  The evolution of the reaction X$_1$ is described by the partial differential equation,
\begin{equation}
\frac{1}{a_1} \left( \frac{\partial n_{\rm X_1}}{\partial t} - K_{zz} \frac{\partial^2 n_{\rm X_1}}{\partial x^2} \right) = {\cal P} - {\cal L} n_{\rm X_1}^{a_1} - {\cal J}_{\rm X_1}.
\end{equation}
The importance of the $1/a_1$ factor cannot be over-stated: it allows for the reaction rates of reactants and products with different stoichiometric coefficients to be placed on the same footing \citep{johnston,steinfeld}.  The production and loss rates are
\begin{equation}
\begin{split}
{\cal P} &= n_{\rm Z_1}^{b_1} n_{\rm Z_2}^{b_2} k_{\rm r}, \\
{\cal L} &= n_{\rm X_2}^{a_2} k_{\rm f}.
\end{split}
\end{equation}
The reaction rate associated with photochemistry is given by ${\cal J}_{\rm X_1}$, which generally depends on $n_{\rm X_1}$.

The diffusion coefficient ($K_{zz}$) is used to \textit{mimic} advection, convection and turbulence and subsume their collective influence into a single free parameter.  Generally, advection, convection and turbulence hardly resemble diffusion in any rigorous way---one often argues that these processes operate on scales that are so small, compared to the characteristic atmospheric length scale of interest, that it ``looks" like diffusion.  The use of $K_{zz}$ is rigorous and exact only for molecular diffusion.  Notwithstanding, the inclusion of a diffusion coefficient allows us to treat situations with disequilibrium chemistry induced by atmospheric motion or mixing without resorting to a full-blown, three-dimensional calculation.

For the product $Z_1$, the evolution equation is
\begin{equation}
\frac{1}{b_1} \left( \frac{\partial n_{\rm Z_1}}{\partial t} - K_{zz} \frac{\partial^2 n_{\rm Z_2}}{\partial x^2} \right) = {\cal P}^\prime - {\cal L}^\prime n_{\rm Z_1}^{b_1} - {\cal J}_{\rm Z_1} ,
\end{equation}
where the production and loss rates are
\begin{equation}
\begin{split}
{\cal P}^\prime &= n_{\rm X_1}^{a_1} n_{\rm X_2}^{a_2} k_{\rm f}, \\
{\cal L}^\prime &= n_{\rm Z_2}^{b_2} k_{\rm r}.
\end{split}
\end{equation}

\subsection{Why photochemistry is a disequilibrium effect}

In the absence of atmospheric mixing ($K_{zz}=0$), we may add the evolution equations for X$_1$ and Z$_1$ to obtain
\begin{equation}
\frac{1}{a_1} \frac{\partial n_{\rm X_1}}{\partial t} + \frac{1}{b_1} \frac{\partial n_{\rm Z_1}}{\partial t} = - {\cal J}_{\rm X_1} - {\cal J}_{\rm Z_1}.
\end{equation}
If we integrate this expression, we obtain
\begin{equation}
\frac{n_{\rm X_1}}{a_1} + \frac{n_{\rm Z_1}}{b_1} = - \int \left( {\cal J}_{\rm X_1} + {\cal J}_{\rm Z_1} \right) dt + {\cal C}.
\end{equation}
If we do the same for all combinations of reactants and products, then we obtain
\begin{equation}
\begin{split}
&\frac{n_{\rm X_1}}{a_1} + \frac{n_{\rm X_2}}{a_2} + \frac{n_{\rm Z_1}}{b_1} + \frac{n_{\rm Z_2}}{b_2} \\
&= - \int \left( {\cal J}_{\rm X_1} + {\cal J}_{\rm X_2} + {\cal J}_{\rm Z_1} + {\cal J}_{\rm Z_2} \right) dt + {\cal C}^\prime,
\end{split}
\end{equation}
where ${\cal C}$ and ${\cal C}^\prime$ are constants of integration.

This result informs us that photochemistry is an intrinsically disequilibrium effect, because it allows the total number of particles in the system to vary with time.  In its absence, the total number of particles is an invariant quantity.

\subsection{Producing chemical equilibrium in the steady-state limit}

If we neglect atmospheric mixing and photochemistry, the steady-state limit of the evolution equations yields
\begin{equation}
n_{\rm X_1}^{a_1} n_{\rm X_2}^{a_2} k_{\rm f} = n_{\rm Z_1}^{b_1} n_{\rm Z_2}^{b_2} k_{\rm r}.
\end{equation}
Since this is identical to the setup in which we used to define our dimensional equilibrium constant ($K^\prime_{\rm eq}$), we conclude that our evolution equations correctly produce chemical equilibrium in the steady-state limit.

\section{Gibbs free energy for hydrogen atom}
\label{append:gibbs}

We use the Gibbs free energy associated with the hydrogen atom from the JANAF database.  Here, we list it in units of kJ/mol/K, from 0 to 6000 K (in increments of 100 K) and at $P_0=1$ bar: 216.035, 212.450, 208.004, 203.186, 198.150, 192.957, 187.640, 182.220, 176.713, 171.132, 165.485, 159.782, 154.028, 148.230, 142.394, 136.522, 130.620, 124.689, 118.734, 112.757, 106.760, 100.744, 94.712	, 88.664, 82.603, 76.530, 70.444, 64.349, 58.243, 52.129, 46.007, 39.877, 33.741, 27.598, 21.449, 15.295, 9.136, 2.973, -3.195, -9.366, -15.541, -21.718, -27.899, -34.082, -40.267, -46.454, -52.643, -58.834, -65.025, -71.218, -77.412, -83.606, -89.801, -95.997, -102.192, -108.389, -114.584, -120.780, -126.976, -133.172, -139.368.  If we denote each of these numbers by $\tilde{G}_{\rm H}$, then we have $\Delta \tilde{G}_0 = -2 \tilde{G}_{\rm H}$ for the net reaction in equation (\ref{eq:hydrogen_reaction}).  It follows that
\begin{equation}
K^\prime = \frac{P}{P_0} \exp{\left( - \frac{\Delta \tilde{G}_0}{{\cal R}_{\rm univ} T} \right)},
\end{equation}
where ${\cal R}_{\rm univ} = 8.3144621$ J K$^{-1}$ mol$^{-1}$ is the universal gas constant.  See \cite{hl15} for more explanation on the unit conversion between $\Delta G_0$ and $\Delta \tilde{G}_0$.

\label{lastpage}


\begin{thebibliography}{99}

\bibitem[Allen \& Yung(1981)]{allen81} Allen, M., \& Yung, Y.L. \ 1981, Journal of Geophysical Research, 86, 3617

\bibitem[Atkins \& de Paula(2006)]{atkins} Atkins, P.W., \& de Paula, J. \ 2006 Physical Chemistry, eighth edition (New York: Freeman)

\bibitem[Barshay \& Lewis(1978)]{barshay78} Barshay, S.S., \& Lewis, J.S. \ 1978, Icarus, 33, 593

\bibitem[Benneke(2015)]{benneke15} Benneke, B. \ 2015, arXiv:1504.07655v1

\bibitem[Blecic et al.(2015)]{blecic15} Blecic, J., Harrington, J., \& Bowman, M.O. \ 2015, arXiv:1505.06392v1

\bibitem[Burrows \& Sharp(1999)]{bs99} Burrows, A., \& Sharp, C.M. \ 1999, ApJ, 512, 843

\bibitem[Ciesla \& Charnley(2006)]{ciesla06} Ciesla, F.J., \& Charnley, S.B. \ 2006, Meteorites and the Early Solar System II, eds. D. S. Lauretta and H. Y. McSween Jr., University of Arizona Press, Tucson, 943 pp., p.209-230

\bibitem[DeVoe(2015)]{devoe} DeVoe, H. \ 2015, Thermodynamics and Chemistry, second edition, sixth version (\texttt{http://www.chem.umd.edu/thermobook}; first edition by Prentice-Hall)

\bibitem[Eisenberg \& Crothers(1979)]{eisenberg} Eisenberg, D., \& Crothers, D. \ 1979, Physical Chemistry with Applications to the Life Sciences (California: Benjamin/Cummings)

\bibitem[Fegley \& Lodders(1996)]{fegley96} Fegley, B., \& Lodders, K. \ 1996, ApJ Letters, 472, L37

\bibitem[Fermi(1936)]{fermi} Fermi, E. \ 1936, Thermodynamics (New York: Dover)

\bibitem[Gail \& Sedlmayr(2014)]{gail} Gail, H.-P., \& Sedlmayr, E. \ 2014 (New York: Cambridge University Press)

\bibitem[Glassman, Yetter \& Glumac(2015)]{glassman} Glassman, I., Yetter, R.A., \& Glumac, N.G. \ 2015, Combustion, fifth edition (Massachusetts: Elsevier)

\bibitem[Heng \& Lyons(2015)]{hl15} Heng, K., \& Lyons, J.R. \ 2015, arXiv:1507.01944v1

\bibitem[Hu, Seager \& Bains(2012)]{hu12} Hu, R., Seager, S., \& Bains, W. \ 2012, ApJ, 761, 166

\bibitem[Hu, Seager \& Bains(2013)]{hu13} Hu, R., Seager, S., \& Bains, W. \ 2013, ApJ, 769, 6

\bibitem[Jacobson(2005)]{jacobson} Jacobson, M.Z. \ 2005, Fundamentals of Atmospheric Modeling (New York: Cambridge University Press)

\bibitem[Johnston(1966)]{johnston} Johnston, H.S. \ 1966, Gas Phase Reaction Rate Theory (New York: Ronald Press Company)

\bibitem[Klotz \& Rosenberg(2008)]{klotz} Klotz, I.M., \& Rosenberg, R.M. \ 2008, Chemical Thermodynamics: Basic Concepts and Methods, seventh edition (New Jersey: Wiley)

\bibitem[Kopparapu, Kasting \& Zahnle(2012)]{koppa12} Kopparapu, R.K., Kasting, J.F., \& Zahnle, K.J. \ 2012, ApJ, 745, 77

\bibitem[Line \& Yung(2013)]{ly13} Line, M.R., \& Yung, Y.L. \ 2013, ApJ, 779, 3

\bibitem[Lodders \& Fegley(2002)]{lodders02} Lodders, K., \& Fegley, B. \ 2002, Icarus, 155, 393

\bibitem[Lodders(2003)]{lodders03} Lodders, K. \ 2003, ApJ, 591, 1220

\bibitem[Madhusudhan(2012)]{madhu12} Madhusudhan, N. \ 2012, ApJ, 758, 36

\bibitem[Moore(1972)]{moore} Moore, W.J. \ 1972, Physical Chemistry, fourth edition (New Jersey: Prentice-Hall)

\bibitem[Moses et al.(2011)]{moses11} Moses, J.I., et al. \ 2011, ApJ, 737, 15

\bibitem[Moses et al.(2013a)]{moses13a} Moses, J.I., Madhusudhan, N., Visscher, C., \& Freedman, R.S. \ 2013a, ApJ, 763, 25

\bibitem[Moses et al.(2013b)]{moses13b} Moses, J.I., et al. \ 2013b, ApJ, 777, 34

\bibitem[Press et al.(2007)]{nr} Press, W.H., Teukolsky, S.A., Vetterling, W.T., \& Flannery, B.P. \ 2007, Numerical Recipes: the Art of Scientific Computing, Third Edition (New York: Cambridge University Press)

\bibitem[Prinn \& Barshay(1977)]{prinn77} Prinn, R.G., \& Barshay, S.S. \ 1977, Science, 198, 1031

\bibitem[Sharp \& Huebner(1990)]{sharp90} Sharp, C.M., \& Huebner, W.F. \ 1990, ApJS, 72, 417

\bibitem[Slater(1939)]{slater} Slater, J.C. \ 1939, Introduction to Chemical Physics (New York: McGraw-Hill)

\bibitem[Smith \& Missen(1982)]{smith} Smith, W.R., \& Missen, R.W. \ 1982, Chemical Reaction Equilibrium Analysis: Theory and Algorithms (New York: Wiley)

\bibitem[Steinfeld, Francisco \& Hase(1989)]{steinfeld} Steinfeld, J.I., Francisco, J.S., \& Hase, W.L. \ 1989, Chemical Kinetics and Dynamics (New Jersey: Prentice-Hall)

\bibitem[Swendsen(2012)]{swendsen} Swendsen, R.H. \ 2012, An Introduction to Statistical Mechanics and Thermodynamics (New York: Oxford University Press)

\bibitem[Upadhyay(2006)]{upa} Upadhyay, S.K. \ 2006, Chemical Kinetics and Reaction Dynamics (Dordrecht: Springer)

\bibitem[van Zeggeren \& Storey(1970)]{zeggeren} van Zeggeren, F., \& Storey, S.H. \ 1970, The Computation of Chemical Equilibria (New York: Cambridge University Press)

\bibitem[Venot et al.(2015)]{venot15} Venot, O., H\'{e}brard, E., Ag\'{u}ndez, M., Decin, L., \& Bounaceur, R. \ 2015, A\&A, 577, A33

\bibitem[Visscher \& Moses(2011)]{vm11} Visscher, C., \& Moses, J.I. \ 2011, ApJ, 738, 72

\bibitem[Yung \& DeMore(1999)]{yung} Yung, Y.L., \& DeMore, W.B. \ 1999, Photochemistry of Planetary Atmospheres (New York: Oxford University Press)

\bibitem[Zahnle et al.(2009)]{zahnle09} Zahnle, K., Marley, M.S., Freedman, R.S., Lodders, K., \& Fortney, J.J. \ 2009, ApJ Letters, 701, L20

\end{thebibliography}
\end{document}